\documentclass[jnl,twocolumn,superscriptaddress]{revtex4}  
\usepackage[draft]{hyperref} 
\usepackage{amsfonts}
\usepackage{graphicx}        
\usepackage{amssymb}
\usepackage{amsmath}
\usepackage{soul}
\usepackage{color}

\newcommand\figref[1]{Fig.~\ref{#1}}

\newcommand{\Psirm}     {\mathrm{\Psi}}

\newcommand{\bfB}   {\mathbf{B}}
\newcommand{\bfH}   {\mathbf{H}}
\newcommand{\bfE}   {\mathbf{E}}
\newcommand{\bfD}   {\mathbf{D}}

\newcommand{\bfk}   {\mathbf{k}}

\newcommand{\bfq}   {\mathbf{q}}

\newcommand{\calM}  {\mathcal{M}}

\begin{document}
	
\title{Homogenization of Layered Media:\\
Intrinsic and Extrinsic Symmetry Breaking\\
 \vskip 4mm
\normalsize\normalfont{Igor Tsukerman$^1$, A N M Shahriyar Hossain$^{1,2}$,
Y. D. Chong$^3$}}
	
\affiliation{Department of Electrical and Computer Engineering,
	  The University of Akron, OH 44325-3904, USA\\
	  igor@uakron.edu\\
	  $^2$Department of Electrical Engineering and Computer Science,
	  Case Western Reserve University, Cleveland, OH, USA 44106
	  \\
  $^3$School of Physical and Mathematical Sciences,
  Nanyang Technological University,
  21 Nanyang Link, Singapore 637371\\
  yidong@ntu.edu.sg
}


\begin{abstract}
A general homogenization procedure for periodic electromagnetic
structures, when applied to layered media with asymmetric lattice cells,
yields an effective tensor with magnetoelectric coupling.
Accurate results for transmission and reflection are obtained
even in cases where classical effective medium theory breaks down.
Magnetoelectric coupling accounts for symmetry breaking in
reflection and transmission when a non-symmetric structure 
is illuminated from two opposite sides.
\end{abstract}


\maketitle


A useful way to understand the properties of a periodic photonic
heterostructure, such as a metamaterial or photonic crystal, is to
represent it as a homogeneous effective medium.  Effective medium
descriptions are known to be accurate in the long-wavelength limit $a
/ \lambda \rightarrow 0$, where $a$ is the unit cell size
	  and $\lambda = 2\pi c / \omega$ is the free-space wavelength, 
	  but break down when
$a / \lambda$ becomes appreciable \cite{Chebykin12,
  Liu-Guenneau-Gralak13, Sheinfux-EMT-breakdown-PRL14,
  Andryieuski-Lavrinenko15, Zhukovsky-EMT-breakdown15, Popov16}.  A
qualitative manifestation of this breakdown occurs when there are
\textit{incompatible symmetries} between the scattering
characteristics of the original heterostructure and the respective
homogenized sample.

\begin{figure}
  \centering
  \includegraphics[width=0.48\textwidth]{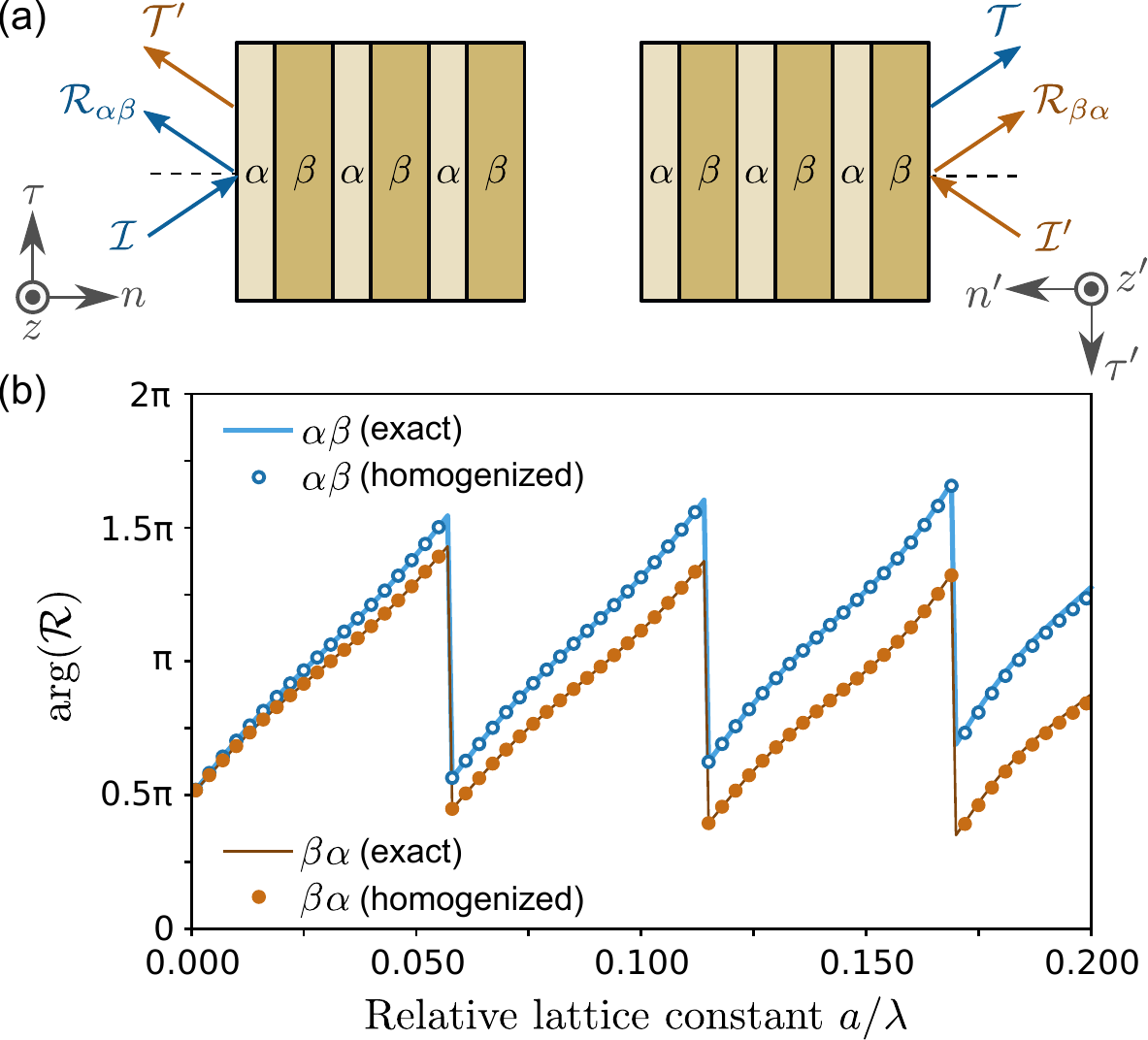}
  \caption{(a) Schematic of a multilayer heterostructure consisting of
    two layers $a$ and $b$ repeated an integer number of times.  The
    layers have dielectric constants $\epsilon_{\alpha}$ and $\epsilon_{\beta}$,
    and are surrounded by air ($\epsilon = 1$).  (b) Phases of the two
    reflection coefficients $R_{\alpha \beta}$ and $R_{\beta \alpha}$ versus 
    relative
    lattice constant $a / \lambda$, where $\lambda$ is the free space
    wavelength.  The multilayer has $\epsilon_{\alpha} = 1$, $\epsilon_{\beta} =
    5$, equal layer widths $d_{\alpha} = d_{\beta} = a/2$, with a total of 5
    lattice periods (10 layers).  The incident plane waves are $s$
    polarized (i.e., $\bfE$ parallel to $z$ and $\bfH$ lying in the
    $n$-$\tau$ plane).  Solid lines: exact values calculated
    via the transfer matrix technique; markers: results for
    the the homogenization method described in the text.}
  \label{fig:layered-medium-setup}
\end{figure}

As an example, consider wave propagation in a dielectric multilayer
consisting of repeated layers labeled $\alpha$ and $\beta$, 
surrounded by air, as shown
in Fig.~\ref{fig:layered-medium-setup}(a).  The layers have unequal
dielectric constants $\epsilon_{\alpha}$ and $\epsilon_{\beta}$ (which may be
complex and dependent on the frequency $\omega$), so that the
heterostructure lacks mirror symmetry with respect to the normal
direction $n$.  In Fig.~\ref{fig:layered-medium-setup}(b), the solid
lines show the phases of the reflection coefficients
$\mathcal{R}_{\alpha \beta}$ and $\mathcal{R}_{\beta \alpha}$, calculated 
analytically
using the transfer matrix technique \cite{Yeh05}, for $s$-polarized waves 
impinging
normally on the structure with $\alpha \beta\ldots$ and $\beta \alpha\ldots$ 
layer orderings, respectively. 
(Reflection coefficients are the ratios of the complex amplitudes of the
electric field in the reflected and incident waves.) 
In the static limit $a / \lambda \rightarrow
0$, the order of the layers is unimportant, but for larger values of
$a/\lambda$ the phases differ substantially
\cite{Sheinfux-EMT-breakdown-PRL14, Andryieuski-Lavrinenko15,
  Zhukovsky-EMT-breakdown15, Lei-EMT-breakdown17}.  We call this
effect, which arises from the lack of mirror asymmetry of the underlying
heterostructure, \textit{intrinsic symmetry breaking} (ISB) --
to be contrasted with \textit{extrinsic} symmetry breaking 
(p.~\pageref{page:ESB}).

At first glance, it seems that ISB cannot be faithfully
reproduced by a homogenized slab, since a homogenous medium would have
an inherent mirror symmetry ensuring that $\mathcal{R}_{\alpha \beta} =
\mathcal{R}_{\beta \alpha}$.  For instance, for $s$-waves and equal
layer widths, standard quasi-static homogenization leads to
a simple dielectric structure with the scalar effective permittivity
$\epsilon_{\mathrm{eff}} = (\epsilon_{\alpha} + \epsilon_{\beta})/2$. 
The symmetry mismatch is also problematic for more sophisticated
homogenization schemes; in particular, it cannot necessarily be
resolved by \textit{nonlocal} effective medium theories
\cite{Chebykin12, Liu-Guenneau-Gralak13, Popov16,
  Mnasri-Rockstuhl-nonlocal18}, whereby effective material parameters 
depend on the Fourier-space wavevector $\bfk$. 
(In real space, this results in non-pointwise relations between
the fields.)  To account for ISB, Lei \textit{et al.}  
introduced an artificial matched layer on the illumination side 
\cite{Lei-EMT-breakdown17}; however,
this is not satisfactory, since an effective medium ought to reflect the
intrinsic characteristics of the structure, independent of the
illumination conditions.

Here, we show that an appropriate \textit{local} homogenization scheme
can accurately account for ISB via magnetoelectric (ME) couplings in
the effective material tensor.  Artificial matching layers
\cite{Lei-EMT-breakdown17} or more complicated nonlocal formulations
are not required.  Details about the homogenization scheme are given
below; when applied to the above multilayer structure, it produces the
values plotted as markers in Fig.~\ref{fig:layered-medium-setup}(b),
which are in excellent agreement with the exact transfer matrix
results.  In particular, $\mathrm{arg}(\mathcal{R}_{\alpha \beta}) =
\mathrm{arg}(\mathcal{R}_{\beta \alpha})$ in the $a/\lambda \rightarrow 0$, 
but these values differ for larger values of $a/\lambda$, so that ISB
is quantitatively accounted for.

A related but different effect arises when the external
media on the two sides of the structure are different.
In that case, not only the phases but also the magnitudes 
of the reflection coefficients may
differ \cite{Sheinfux-EMT-breakdown-PRL14}.  We refer to this as 
\textit{extrinsic symmetry breaking} (ESB)\label{page:ESB}.  When the medium
on one side is optically denser than the static average in the slab,
different layer orderings can produce extremely different reflection
coefficients when the incident wave is close to the critical angle for
total internal reflection (the phenomenon does not contradict optical
reciprocity \cite{Potton-reciprocity04}).  As we shall see, the
effects of ESB can also be successfully accounted for by
homogenization with ME coupling.

To understand the origin of ME coupling, we
first consider the simple case of a lossless homogeneous medium with
two $s$-polarized plane waves propagating in opposite axial directions
$\pm n$, indicated in Fig.~\ref{fig:layered-medium-setup}(a).  The respective
equal-magnitude wave vectors are $\pm \bfq = \pm k_0
\hat{\mathbf{q}}$, where $\hat{\mathbf{q}}$ is a unit vector,
and the respective electric and magnetic field components are
\begin{equation}\label{eqn:two-plane-waves-EH}
  {E}_{z}^\pm = {E}_{0}^\pm
  \exp(\pm i k_0 \hat{\mathbf{q}} \cdot \mathbf{r}),
   ~~
  {H}_{\tau}^\pm = H_{0}^\pm
  \exp(\pm i k_0 \hat{\mathbf{q}} \cdot \mathbf{r}),
\end{equation}
where $\mathbf{r} = (n, \tau, z)$.  The waves satisfy Maxwell's
equations $\nabla \times \bfE = i k_0 \bfB$ and $\nabla \times \bfH =
-i k_0 \bfD$ under the $\exp(-i \omega t)$ phasor convention.  The
medium is described by a material tensor $\calM$ whose matrix
representation $M$ in a given coordinate system satisfies
\begin{align}
  \label{eqn:DB-eq-M-EH-generic}
  \Psirm_{DB} &= M \, \Psirm_{EH}, \\
  \label{eqn:DB-EH-2plane-waves}
  \Psirm_{DB} &=
  \begin{bmatrix} 
    D_{0z}^+ & D_{0z}^- \\ 
    B_{0 \tau}^+ & B_{0\tau}^- 
  \end{bmatrix}, \;\;
  \Psirm_{EH} =
  \begin{bmatrix} 
    E_{0z}^+ & E_{0z}^-\\ 
    H_{0\tau}^+ & H_{0\tau}^- 
  \end{bmatrix}.
\end{align}
Normalizing the $E_{0z}^\pm$ amplitudes to unity and applying
Maxwell's equations gives
\begin{equation}\label{DB-EH-generic-normalized}
  \Psirm_{DB} =  q_n
  \begin{bmatrix} -Y_+ & Y_-\\ -1  & 1 \end{bmatrix},
  \quad
  \Psirm_{EH} = \begin{bmatrix} 1 & 1\\ Y_+ & Y_- \end{bmatrix},
\end{equation}    
where $Y_{\pm} = H_{0\tau}^\pm / E_{0 z}^\pm$ are the wave admittances.  Then 
\begin{align}
  \label{eqn:eff-tensor-via-admittances-generic}
  M = \Psirm_{EH}^{-1} \Psirm_{DB}
  = \frac{q_n} {Y_2 - Y_1}
  \begin{bmatrix}
    -2 Y_1 Y_2 & Y_1 + Y_2\\
    -(Y_1 + Y_2)   &  2
  \end{bmatrix}.
\end{align}
The ME coupling is represented by the off-diagonal terms in
$M$, and is absent if and only if $Y_1 = -Y_2$.  This is the
case for an ordinary homogeneous dielectric medium.

Note that this ME coupling is not equivalent to optical activity.  It
does not alter the polarization of the wave, which remains 
$s$-polarized throughout.  By contrast, the constitutive relations for
optically active media typically include a contribution to $\bfD$ in
the direction of $\bfB$, as well as a contribution to $\bfB$ in the
direction of $\bfE$
\cite{Bohren-active-sphere74,Bassiri-Papas-Engheta88,Engheta-Jaggard88}.

We now adopt the homogenization scheme described in
Refs.~\onlinecite{Tsukerman-Markel14, Tsukerman-PLA17}, which
generates an effective tensor $\calM$ for a layered heterostructure by
approximating the fields on two scales, one finer and the other one
coarser than the lattice cell size.  The fine-scale fields are
approximated by basis sets of Bloch waves traveling in different
directions.  The coarse-scale fields consist of the respective
generalized plane waves, which must satisfy (i) Maxwell's equations
within the sample and (ii) Maxwell's boundary conditions for the
tangential components of the electric and magnetic fields on the
boundary of the sample.  To satisfy (ii), the plane wave amplitudes $\bfE_{0
  \alpha}$ and $\bfH_{0 \alpha}$ are computed as boundary averages of
the periodic factors of Bloch waves; to satisfy (i), the $M$ matrix
is found by solving a linear algebra problem analogous to
\eqref{eqn:DB-eq-M-EH-generic}, except that now the $\Psirm_{DB}$
and $\Psirm_{EH}$ matrices are rectangular, with the number of columns
equal to the number of basis functions. Hence \eqref{eqn:DB-eq-M-EH-generic}
is in general interpreted in the least squares sense 
rather than as an exact equality
\cite{Tsukerman-Markel14,Tsukerman-PLA17}.

We apply this procedure to $s$-polarized waves 
in the multilayer heterostructure of Fig.~\ref{fig:layered-medium-setup}(a) 
(the $p$-wave case can be dealt with similarly). To get an analytical
insight, we take all the materials to be lossless, so that $\epsilon_{\alpha}$ 
and
$\epsilon_{\beta}$ are real, and consider a fine-scale basis of only two Bloch 
waves,
with their Bloch wave-numbers $\pm q_n$ at the operating frequency:
\begin{align}
  e_1(n) &= u(n) \exp( iq_n n) \label{eqn:Bloch-wave1-e} \\
  h_{1 \tau}(n) &= 
  -k_0^{-1} [q_n u(n) - i u'(n) ] \exp(iq_n n)
  \label{eqn:Bloch-wave1-h} \\
  e_2(n) &= u^*(n) \exp(-iq_n n) \label{eqn:Bloch-wave2-e} \\
  h_{2 \tau}(n) &=
  k_0^{-1} [q_n u^*(n) + i u^{*\prime} (n) ]  \exp(-iq_n n).
  \label{eqn:Bloch-wave2-h}
\end{align}
Here $u(n)$ denotes the lattice-periodic factor for the
electric field, and `*' indicates complex conjugates. 
On the coarse scale, our procedure defines the EH-amplitudes
of plane waves as the \textit{boundary values} ($n = 0$) of 
the Bloch waves:
\begin{align}
  E_{0z}^+ &= u(0), &
  H_{0 \tau}^+ &=  - k_0^{-1} [q_n E_{0}^+ - i u'(0)],  \\
  E_{0z}^- &= u^*(0), &
  H_{0 \tau}^- &= \;\;\,
  k_0^{-1} [q_n E_0^- + i u^{*\prime}(0) ].
  \label{eqn:PW2-H0}
\end{align}
The corresponding admittances are therefore
\begin{align}
  Y_1  &= -k_0^{-1} \big[q_n  - i \tilde{e}_1'(0)\, / \tilde{e}_1(0) \big]
  \label{eqn:admittance-1} \\ 
  Y_2 &= \;\;\,
  k_0^{-1} \big[q_n  + i \tilde{e}_1^{*\prime}(0) / \tilde{e}_1^{*}(0) \big].
  \label{eqn:admittance-2}
\end{align}
An explicit expression for the ME coupling term is
\begin{equation}\label{eqn:ME-term}
  M_{12} = i \, \frac{q_n}{k_0} \, \frac{\mathrm{Re} \,\eta}
  {\mathrm{Im} \, \eta + q},
  \quad
  \eta \equiv \frac{u'(0)}{u(0)}.
\end{equation}
Hence the ME coupling arises from the difference in the boundary
admittances of the Bloch waves.  If the lattice cell possesses mirror
symmetry, then derivative $u'(0)$ vanishes, so $M_{12} = 0$.

Importantly, the matrix representation of $\calM$ depends on the choice
of coordinate system.  If $(n, \tau, z)$ is switched to the
mirror-image system $(n', \tau', z)$ shown in
Fig.~\ref{fig:layered-medium-setup}(a), the off-diagonal ME terms
reverse sign.  The effective medium is thus able to capture the
effects of the broken mirror symmetry, as demonstrated by the numerical
results of Fig.~\ref{fig:layered-medium-setup}(b).

\begin{figure}
  \centering
  \includegraphics[width=0.48\textwidth]{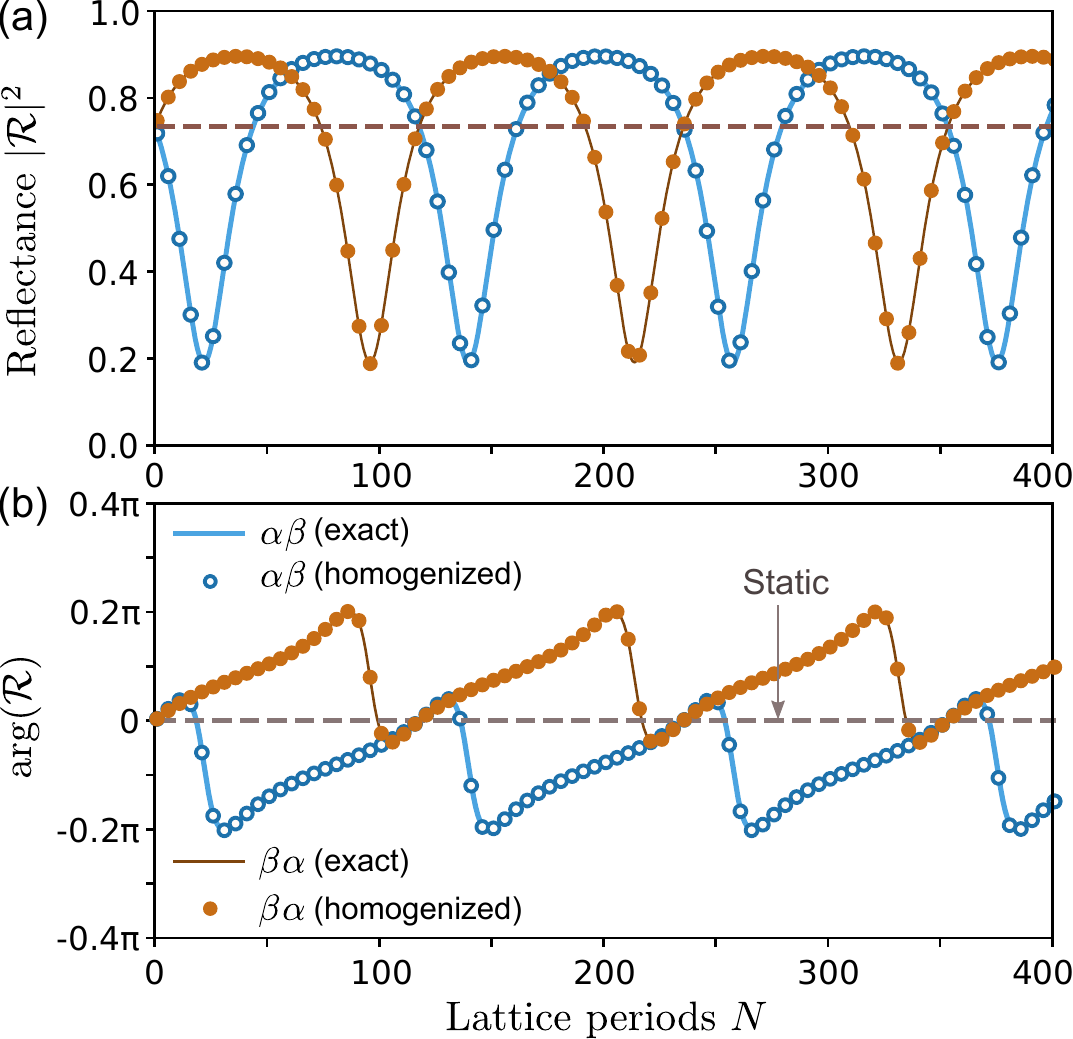}
  \caption{Scattering characteristics of a multilayer with asymmetric
    external media. The structure consists of $N$ identical cells,
    each containing two layers, `$\alpha$' and `$\beta$', with the thicknesses 
    $d_{\alpha} = d_{\beta} = 10$~nm and dielectric
    constants $\epsilon_{\alpha} = 5$ and $\epsilon_{\beta} = 1$. Free-space
    wavelength $\lambda = 500$~nm (i.e. $a/\lambda = 0.04$).
    External dielectric constants are $\epsilon_{\mathrm{in}} = 4$
    and $\epsilon_{\mathrm{out}} = 3$ on the sides of incidence and 
    transmission, respectively. (These parameters match the ones
    of Ref.~\onlinecite{Sheinfux-EMT-breakdown-PRL14}.)
    Incidence is at the critical angle \eqref{eqn:theta-crit}.
    (a) reflectances $|\mathcal{R}_{\alpha \beta}|^2$ 
    and $|\mathcal{R}_{\alpha \beta}|^2$; (b) reflection coefficient phases
    $\mathrm{arg}(\mathcal{R}_{\alpha \beta})$ and
    $\mathrm{arg}(\mathcal{R}_{\alpha \beta})$ vs. $N$.  
    Solid curves: exact values calculated via the transfer 
    matrix method. Markers: results for homogenization with
    magnetoelectric coupling 
    \cite{Tsukerman-Markel14,Tsukerman-PLA17}. 
    Two layer orderings, $\alpha \beta\dots$ and $\beta \alpha\dots$,
    exhibit significantly different behaviors. Static
    homogenization (horizontal dashed lines) yields
qualitatively inaccurate results.}
  \label{fig:R-vs-numcells-Sheinfux-PRL2014}
\end{figure}

Turning now to ESB effects, we set the same parameters as in
\cite{Sheinfux-EMT-breakdown-PRL14}:
equal layer widths $d_{\alpha} = d_{\beta} = a/2 = 10$~nm,
dielectric constants $\epsilon_{\alpha} = 5$ and $\epsilon_{\beta} = 1$, and
free-space wavelength $\lambda = 500\,\textrm{nm}$.  The effective
permittivity for the $s$ mode in the static limit ($a / \lambda
\rightarrow 0$) is $\epsilon_{\mathrm{stat}} = (\epsilon_{\alpha} d_{\alpha} +
\epsilon_{\beta} d_{\beta})/a = 3$. The number of lattice cells $N$ is allowed 
to vary.
External dielectric permittivities are $\epsilon_{\mathrm{in}} = 4$
and $\epsilon_{\mathrm{out}} = 3$ on the sides of incidence and 
transmission, respectively. These parameters match the ones
of Ref.~\onlinecite{Sheinfux-EMT-breakdown-PRL14}
and, since $\epsilon_{\mathrm{in}} \neq \epsilon_{\mathrm{out}}$,
give rise to ESB.
In the static limit, the critical angle for total internal reflection is
\begin{equation}\label{eqn:theta-crit}
  \theta_{\mathrm{crit}} = \sin^{-1} \sqrt{\epsilon_{\mathrm{stat}} /
    \epsilon_{\mathrm{in}}} = 60^{\circ}.
\end{equation}
Once the angle of incidence reaches the critical value, the wave in
the $b$ layer becomes evanescent, but the wave in the $a$ layer is
propagating.
%
%
%
Near $\theta_{\mathrm{crit}}$, the reflection and transmission
coefficients are found to depend strongly on the choice of layer order
($\alpha \beta\dots$ or $\beta \alpha\dots$), and both are very different from 
the
value predicted by the static permittivity $\epsilon_{\mathrm{stat}}$;
see \cite{Sheinfux-EMT-breakdown-PRL14} and
\figref{fig:R-vs-numcells-Sheinfux-PRL2014}.

\begin{figure}
  \centering
  \includegraphics[width=0.48\textwidth]{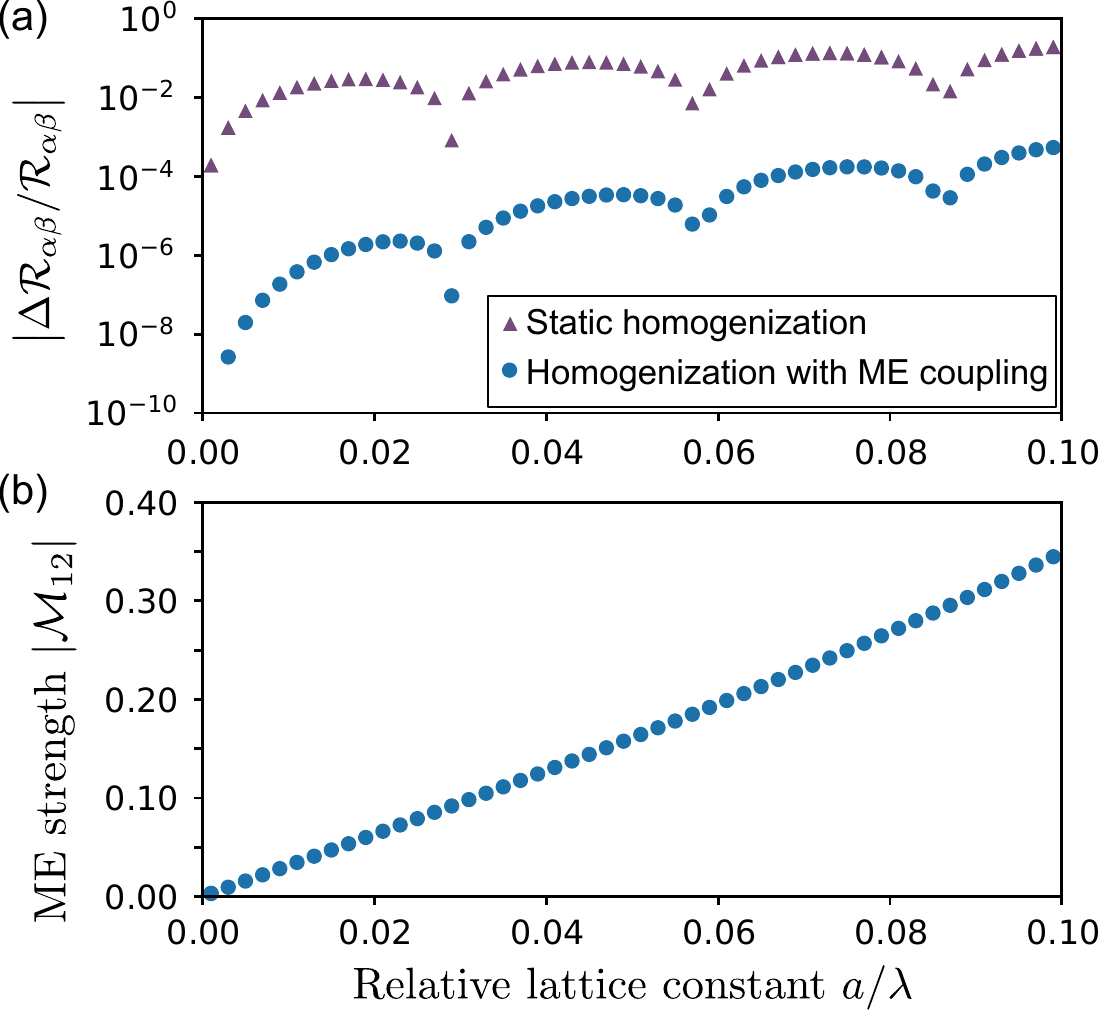}
  \caption{ Relative errors and effective ME coupling strengths versus
    normalized lattice constant $a/\lambda$, with all other parameters the same
    as in Fig.~\ref{fig:R-vs-numcells-Sheinfux-PRL2014}.  (a) Relative
    error in $\mathcal{R}_{\alpha \beta}$, calculated using the static and ME
    homogenization schemes and compared to the transfer matrix
    results.  Note the logarithmic scale on the vertical axis.  (b)
    Magnitude of the off-diagonal element in the effective
    $M$ matrix, which describes the strength of the ME
    coupling.}
  \label{fig:homog-errors}
\end{figure}

For this setup, the fine-scale basis in our homogenization procedure 
contains Bloch waves with the tangential components of the Bloch
wave vector $q_{\tau m} = m n_{\mathrm{stat}} k_0 /(n_q - 1)$,
where $0 \le m < n_q$.  Only non-negative values of $q_{\tau}$ are
needed due to the symmetry of the structure in the tangential
direction. For each $q_{\tau}$ there are two Bloch waves in the
forward and backward directions; hence
the total size of the Bloch basis is $2n_q$. On the coarse level, the
basis consists of the respective $2n_q$ generalized plane waves.  We
will take $n_q = 7$; the results shown below are essentially unchanged
for other choices of $n_q \geq 5$.

Figure~\ref{fig:R-vs-numcells-Sheinfux-PRL2014} plots the reflectance,
and the phase of the reflection coefficient
against the number of lattice periods $N$ in the
slab. (Each until cell has two layers, so the total number of layers
is $2N$.)
%
%
For all quantities, the results of the homogenization
scheme agrees extremely well with the exact results obtained by
transfer matrix calculations.  As an example,
Fig.~\ref{fig:homog-errors}(a) shows that the relative error in the
reflection coefficient $\mathcal{R}_{\alpha \beta}$ is at least two orders of
magnitude lower than in the static approximation.  Notably, the
homogenization captures the substantial differences between the
$\alpha \beta\dots$ and $\beta \alpha\dots$ layer orderings 
\cite{Sheinfux-EMT-breakdown-PRL14}
via the ME coupling in the $\mathcal{M}$ tensor. For instance,
\begin{equation}\label{eqn-material-tensor-Sheinfux-PRL2014-004}
   M(a/\lambda = 0.04) ~\approx
\begin{bmatrix}
3.02	 & 0 &	–0.128i	\\
0 &	1	& 0	\\
0.128i	& 0 &	1
\end{bmatrix}.
\end{equation}
The effective permittivity differs slightly from its static value of
3, but the key feature is the presence of the ME coupling terms, which
are clearly appreciable.  The magnitude of the ME coupling is
approximately proportional to $a/\lambda$ and vanishes in the static
limit $a/\lambda \rightarrow 0$, as shown in
Fig.~\ref{fig:homog-errors}(b).

In conclusion, we have demonstrated that a \textit{local}
homogenization procedure, which produces an effective material
tensor with magnetoelectric coupling terms, 
can accurately describe the behavior of periodic
multilayer heterostructures away from the static limit.  Specifically,
the local homogenization correctly accounts for the effects of
intrinsic and extrinsic symmetry breaking, one manifestation
of which is a dependence of the reflection and transmission 
characteristics on the order of the layers, as noted in previous studies
\cite{Sheinfux-EMT-breakdown-PRL14, Lei-EMT-breakdown17}.  

\vskip 2mm

The research of IT and ANMSH was supported in part by the US National
Science Foundation awards DMS-1216970 and DMS-1620112.  CYD was
supported by the Singapore MOE Academic Research Fund Tier 3 Grant
MOE2016-T3-1-006.  

\medskip



%

\bigskip

\bibliography{magnetoelectric_coupling_Letter}




\end{document}